# A new compressive video sensing framework for mobile broadcast

Chengbo Li, Hong Jiang and Paul Wilford and Yin Zhang and Mike Scheutzow

*Abstract*—A new video coding method based on compressive sampling is proposed. In this method, a video is coded using compressive measurements on video cubes. Video reconstruction is performed by minimization of total variation (TV) of the pixelwise DCT coefficients along the temporal direction. A new reconstruction algorithm is developed from TVAL3, an efficient TV minimization algorithm based on the alternating minimization and augmented Lagrangian methods. Video coding with this method is inherently scalable, and has applications in mobile broadcast.

## I. INTRODUCTION

In a mobile video broadcast network, a video source is broadcast to multiple clients having different characteristics. The clients may have different channel capacities, different display resolutions, or different computing resources. It is therefore desirable for a video source to be encoded in such a way that a single encoded video stream can be transmitted, and be optimally usable by all the diverse clients. In other words, we want to encode and transmit the video source once, but allow any subset of the bit stream to be successfully decoded by a client, possibly at different complexities.

The traditional video coding such as MPEG (MPEG2 or MPEG4) does not provide the scalability described above. The lack of scalability exhibits itself in at least two ways. First, an encoded video is not scalable with transmission channel capacity. Because of its fixed bit rate, an encoded stream is unusable in a channel which does not support the required bit rate, and is also suboptimal in a channel having higher bit rate. For MPEG coders, uncorrected transmission errors cause reduced video quality, and the rapid drop off in quality as error rate increases is known as the "cliff effect." Second, the MPEG video is not scalable with decoding complexity or decoded quality. An encoded video bitstream decodes only one way and with a fixed complexity (not considering post-processing such as resizing, or enhancement, after decoding). It is necessary to create multiple encoded streams of the same video content to support decoders of different qualities.

Efforts have been made to introduce multi-resolution scalability into video coding, noticeably by the scalable video coding (SVC) feature of H.264 [1] and the wavelet transform of Motion JPEG 2000 [2]. Both methods encode video into ordered layers, or levels, of streams, and the resolution, or quality, of the decoded video increases progressively as higher layers are processed at the decoder. Hierarchical modulation [3] may be used in conjunction with these scalable video codes to achieve more bandwidth efficiency. For example, the high priority of hierarchical modulation can be used to carry lower layers of the encoded video, and low priority of hierarchical modulation can be used to carry higher layers of the encoded video.

There has been an abundance of research activity in video coding to provide scalable decoding resolution, see [4-7]. A joint video coding and transmission method was proposed in [8] to provide scalability with transmission channel capacity. These activities are a response to the fact that the scalability provided by H.264 or Motion JPEG 2000 is still not satisfactory. Specifically, the ordered layer structure does not provide scalability at a fundamental level, because a video encoded using these approaches must be decoded at the lowest layer, and progressively built up by higher layers. The loss of a lower layer in transmission makes the higher layer data useless, even when the higher layer data is received error-free. Therefore, the ordered layer structure is not scalable with the channel capacity [8]. Protecting an encoded (non-scalable) video stream using fountain codes does not solve this problem, because a fountain code only provides reliable delivery of the encoded video stream, but does not provide scalability to the video itself. Other error-resilient video transmission methods are considered in [24-25]

Based on early efforts on $\ell_1$ minimization, Candès, Romgerg, Tao [9, 19], and Donoho [20] developed compressive sensing (CS) theory which proves that a sparse signal under some sparsity basis can be precisely recovered using a small number of measurements. Due to the proliferation of compressive sampling techniques, video coding using compressive measurements is rapidly emerging [10-11]. Compressive video sensing offers the scalability desired in a video network [12-13], and is also suitable for lossy wireless transmission [14]. When measurements of a video are made using a random (or pseudo-random) matrix, the video source information is distributed among the measurements of equal significance. Particularly, this means that no one measurement is more important than another. The reconstruction of video requires a certain number of measurements to be available, but it does not require the availability of a particular measurement. In this sense, a measurement lost in transmission can simply be replaced by any other measurement. Furthermore, since a video does not have a well-defined sparsity, statistically, when more measurements are used in reconstruction, the image quality of the reconstructed video improves [15]. If the measurements of the video are transmitted by broadcast or multicast, a receiver in a higher capacity channel will have more measurements available, and hence a reconstructed video of higher quality, than a receiver in a lower capacity channel. These properties

Chengbo Li and Yin Zhang are with Computational and Applied Mathematics department, Rice University.
Hong Jiang, Paul Wilford and Mike Scheutzow are with Bell Laboratories, Alcatel-Lucent.

illustrate that video coding using compressive sampling is inherently scalable with the channel capacity, and does not suffer from the cliff effect in broadcast or multicast.

In this paper, we propose a framework for video coding using compressive measurements in which an encoded video is scalable both with the channel capacity and with decoding complexity. In this framework, a source video is divided into self-contained video cubes. A random matrix is used for making the measurements. The resulting measurements are indistinguishable in the sense that the quality of the reconstructed video improves as the number of measurements increase, but does not rely on any particular set of measurements. In the proposed framework, the reconstruction is performed by minimizing the total variation (TV) of the coefficients of the pixelwise discrete cosine transform (DCT) in the temporal direction. TV minimization was first introduced for image denoising problems by Rudin, Osher, and Fatemi [21], and applied to compressive sensing by Candès, Romgerg, Tao [9]. In general, it succeeds when the gradient of the underlying signal or image is sparse. However, TV has been shown to be effective for image reconstruction on a much larger scale empirically. The discrete cosine transform has been widely used as the sparsity basis for compressive sensing due to its ability to remove the periodic redundancy in a signal. Therefore, the proposed TV-DCT method has the advantage of preserving the spatial edges of images using TV, and, at the same time, removing the temporal redundancy using the DCT.

Also in this paper, we extend the TVAL3 algorithm [17], originally developed for solving 2D TV minimization problems, to the application of video reconstruction. A new algorithm based on TVAL3 is developed to solve the TV-DCT minimization problem of this paper.

The focus of this paper is on using compressive sensing as an effective coding method for video transmission in broadcast channels. In particular, we address the issue of how to avoid the cliff effect and how to encode the video so that a single encoded bit stream can be used in channels of different bandwidth, and by decoders of different resolution. In this paper, this goal is achieved by using compressive measurements on video cubes. During video reconstruction, the spatial total variation (TV) and temporal discrete cosine transform (DCT) are used as sparsifying operators.

The method of this paper is similar to the methods used in MRI imaging [27-28] in the sense that the signals, video in this paper, and dynamic MRI images in [27-28], are sparse both spatially and temporally, and the sparsity can be exploited by using spatial and temporal transforms. The MRI methods, however, are different. In [27], the measurements are made in the Fourier transform domain while the measurements of video in this paper are made in the time domain. The method of [28] uses predictions and residuals similar to MPEG, so it suffers from the cliff effect.

The method of this paper is also different from that of [13] in two aspects. First, key frames are used in [13]. The use of key frames causes cliff effects because the loss of a key frame has a significant impact on the decoded video. Secondly, a feedback channel must be used in [13] while we are concerned with broadcast in which a feedback channel is typically not available.

The paper is organized as follows. The new framework for video coding is introduced in Section II. A new algorithm extended from TVAL3 is described in Section III. Numerical results are presented in Section IV. The conclusion and discussion are covered in Section V.

## II. VIDEO CODING USING COMPRESSIVE SAMPLING

In this section, a framework is developed in which a source video is coded using compressive sampling, broadcast in a wireless network and reconstructed by multiple clients as shown in Figure 1.

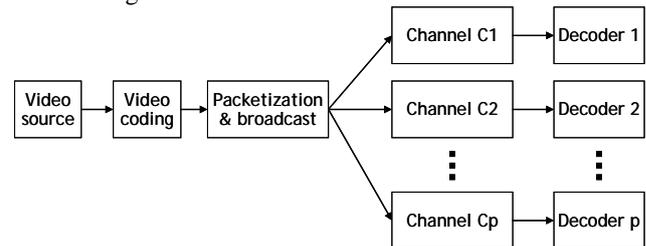

**Figure 1. Video coding for wireless transmission**

Our goal is to use an efficient video coding that is scalable in both capacity and decoding complexity. In this framework, the source video is first divided into video cubes. Then the measurements of each cube are generated by random linear combinations of pixels and transmitted. In the decoder, video reconstruction is performed by minimizing the two dimensional TV of the time domain DCT coefficients for each video cube.

### A. Video coding using compressive sampling

A source video consists of a number of frames with a resolution of $P \times Q$, where $P$ and $Q$ are the numbers of the horizontal and vertical pixels, respectively, for each frame of the video. To efficiently encode it, we process the source video block by block. First, the video source is divided into non-intersecting cubes, each of which consists of $r$ frames of a sub-region of size $p \times q$. For simplicity of description, we assume that each frame of a video cube is taken from the same spatial region in its respective frame of the source video. A video cube is illustrated in Figure 2 in which the video cube has 4 frames, and each is taken from a sub-region in the top right corner of a full video frame.

The encoding of video is performed cube by cube on all video cubes using compressive sensing. Let $x \in R^n$ be the column vector obtained by concatenating the pixels of a video cube column by column, and then frame by frame, i.e., $x$ is resulted from the vectorization of the 3-D video cube, where $n = p \times q \times r$ is the length of $x$. In general, the pixels in a video cube, especially when the frames of the cube are chosen by a motion estimate or motion compensation scheme, are highly correlated, and therefore, the vector $x$ or the gradient of $x$ is sparse under some basis. This means that $x$ can be well represented and reconstructed by using compressive sensing [9]. Suppose $A \in \Re^{m \times n}$ denotes the measurement matrix ($m \leq n$), then the $m$ compressive measurements of $x$ compose the vector $y \in \Re^m$. Mathematically, we have

$$y = Ax. \qquad (2.1)$$

The encoding process described above is illustrated in Figure 2.

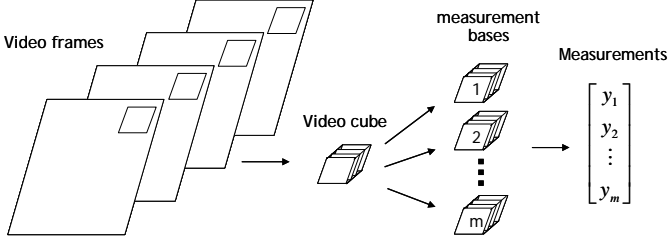

**Figure 2. Video encoding using compressive sensing**

The measurement matrix $A$ should be incoherent with the sparsity basis of signal $x$, but in general, a matrix generated by randomly selecting the entries of the columns can result in good performance [20]. A matrix built by randomly picking rows from the Fourier matrix has been proven to possess a strong recoverability [9]. In this paper, a randomly permutated Walsh-Hadamard matrix is used. The Walsh-Hadamard matrix is defined by a recursion by using the Kronecker product. This class of matrices has the advantage of easy implementation on hardware, and satisfactory recoverability. In practice, the measurements are made by using a fast transform, so there is no need to generate the Walsh-Hadamard matrix explicitly.

The measurements from a random matrix have equal significance in the sense that the quality of the reconstructed video does not depend on any particular, or a particular set of, measurements. The quality of the reconstructed video is a function of the number of distinct measurements received. This property is particularly desirable in wireless applications. In wireless broadcast, many measurements are transmitted, but different clients may correctly receive different subsets of measurements due to different channel conditions. A client with a higher channel capacity receives more correct measurements, and therefore, is able to reconstruct a video with a higher quality than a client with a lower channel capacity. This provides a graceful degradation in the broadcast system, and avoids the cliff effect.

This approach has an advantage even in a non-broadcast network. When the compressive measurements are used for video transmission between two stations, the receiving station does not need to acknowledge whether a measurement is received or not. The sending station only needs to keep transmitting (distinct) measurements until the receiving station acknowledges that a sufficient number of measurements have been received, after which the transmission could stop. Since only one acknowledgement is sent, it reduces overhead and latency in the transmission.

It is also worthwhile to note that each measurement $y_i$ has a distribution approximating the Gaussian distribution. This is because if the number of pixels in a video cube is sufficiently large and the entries of the measurement matrix are independent random variables, the central limit theorem applies to each measurement which is a sum of independent random variables. Knowledge of the distribution of the measurements helps to make more efficient quantization of measurements.

### B. Reconstruction

The received measurements $y$ may contain noise due to the poor channel condition or signal disturbance in the course of wireless transmission. We decode the video by reconstructing each of the video cubes. A video cube is recovered from the received measurements by solving either a constraint minimization problem

$$\min_x \Phi(x) \text{ such that } y = Ax, \qquad (2.2)$$

for a noise-free case, or an unconstrained problem

$$\min_x \Phi(x) + \frac{\mu}{2}\|Ax - y\|_2^2, \qquad (2.3)$$

for a noisy case. Here, $\mu$ is the penalty parameter to balance two parts and $\Phi(x)$ is the regularization term to handle the ill-posedness or to prohibit overfitting.

Usually, there are three types of the regularization term, namely, $\ell_0$-norm which counts the number of non-zero entries, $\ell_1$-norm which sums up the magnitude of all entries, and TV-norm which measures the discontinuities in a signal. The $\ell_0$ minimization is unstable and has exponential complexity. There is extensive literature illustrating results on equivalence of $\ell_0$ and $\ell_1$ minimization for the problem (2.2) or (2.3), but solving $\ell_1$ minimization requires much lower complexity [18, 19]. However, when $x$ is the vector formed from the pixels of a video cube, it is not obvious in which basis $x$ is sparse, and furthermore, in which basis, $x$ has the most sparseness.

The total variation regularization has been widely, and successfully, applied to image processing [15, 17, 21, 23]. It is capable of detecting edges and sharp changes in a signal or 2D image. It has been shown recently that TV transform is equivalent to wavelet transform in image restoration [29]. We propose a new regularization based on total variation and suitable for the video cube reconstruction.

To define the regularization term, we reshape the vector $x$ into a 2D matrix. Each column of $x$ is formed by pixels of a frame, and the rows of $x$ are formed by the frames in the video cube in the temporal direction. The new regularization term calculates the spatial TV of the time domain DCT coefficients of current cube, i.e.,

$$\Phi(x) = TV_s \cdot x \cdot DCT_t. \qquad (2.4)$$

Specifically, each column of $TV_s \cdot Z$ is the framewise 2D total variation of 3D cube $Z$ defined as

$$TV_s \cdot Z = \sum_{i,j,k} \sqrt{\left(Z_{i+1,j,k} - Z_{i,j,k}\right)^2 + \left(Z_{i,j+1,k} - Z_{i,j,k}\right)^2},$$

for isotropic total variation, or

$$TV_s \cdot Z = \sum_{i,j,k}(|Z_{i+1,j,k} - Z_{i,j,k}| + |Z_{i,j+1,k} - Z_{i,j,k}|),$$

for anisotropic total variation. For vector $z$, generated by stacking columns of $Z$, $TV_s \cdot z$ is defined to be equal to $TV_s \cdot Z$, i.e.,

$$TV_s \cdot z = \sum_{l=1}^{n} \|D_l z\|. \quad (2.5)$$

In (2.5), $D_l z \in \Re^2$ is the discrete gradient of $Z$ at position $l$, given by

$$D_l z = (Z_{i+1,j,k} - Z_{i,j,k}, Z_{i,j+1,k} - Z_{i,j,k})^T, \quad (2.6)$$

where index $l$ is the 1D correspondence of the 3D index $(i,j,k)$. Also in (2.5) $\|\cdot\|$ can be either the $\ell_1$-norm which corresponds to the anisotropic TV or the $\ell_2$-norm which corresponds to the isotropic TV.

In (2.4), $x \cdot DCT_t$ represents the temporal discrete cosine transform of each row of $x$. Each column of $x \cdot DCT_t$ consists of DCT coefficients of a particular frequency. In short, using the TV-DCT regularization as the objective function minimizes the spatial total variation of the frequency components in time. The new regularization defined above is illustrated in Figure 3.

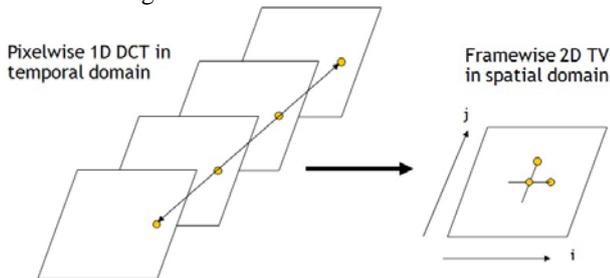

**Figure 3. Illustration of TV-DCT regularization**

In Figure 3, the left hand side illustrates a source video cube with 4 frames. For each spatial location, pixels in the same spatial location from all frames are used in a 1D DCT transform. After the temporal DCT transform, each frame consists of DCT coefficients of a particular frequency. On the right hand side of Figure 3, the DCT coefficients from each frame are used to perform the 2D spatial TV calculation of equation (2.5).

## III. EXTENDED TVAL3 ALGORITHM

In this section, we develop a new algorithm which is extended from TVAL3 [17] for the solution of problems (2.2) and (2.3). TVAL3 is a TV minimization algorithm based on the augmented Lagrangian method [22] and alternating minimization, which is widely used to solve 1D signal and 2D image reconstruction and denoising problems. The idea of alternating minimization was introduced to solve a series of TV minimization problems including deconvolution, denoising, and image reconstruction by Zhang *et al.* [23].

At each outer iteration of TVAL3, the algorithm minimizes the corresponding augmented Lagrangian function by solving two subproblems alternately. One subproblem is separable and has a closed-form solution while the other subproblem is quadratic and can be approximated by one-step steepest descent with an aggressive step length. We can employ similar strategy to solve both (2.2) and (2.3). Since the derivation of the algorithm is almost the same for both problems, in the following, we will present the description of the algorithm for (2.2) in details. The algorithm for (2.3) is similar.

Let $T_i \in \Re^{2 \times n}$ denote the linear transformation satisfying $T_i x = D_i(x \cdot DCT_t)$ for any $i$, $1 \leq i \leq n$, where the operator $D_i$ is defined in (2.6). Then the minimization problem (2.2) becomes

$$\min_x \sum_{i=1}^{n} \|T_i x\| \text{ such that } y = Ax. \quad (3.1)$$

By introducing a series of new variables $w_i = T_i x$, $1 \leq i \leq n$, we consider a variant of (3.1)

$$\min_{x,w_i} \sum_{i=1}^{n} \|w_i\| \text{ s.t. } y = Ax \text{ and } w_i = T_i x, \text{ for all } i. \quad (3.2)$$

Its corresponding augmented Lagrangian function is

$$L_A(x, w_i) = \sum_{i=1}^{n} (\|w_i\| - v_i^T(T_i x - w_i) + \frac{\beta_i}{2}\|T_i x - w_i\|_2^2) - \lambda^T(Ax - y) + \frac{\mu}{2}\|Ax - y\|_2^2. \quad (3.3)$$

Let $x^*$ and $w_i^*$ denote the true minimizers of $L_A(x, w_i)$. The augmented Lagrangian method minimizes (3.3) recursively and then updates multipliers $v_i$ and $\lambda$ at the end of each iteration as follows:

$$\begin{cases} \tilde{v}_i = v_i - \beta_i(T_i x^* - w_i^*) & \text{for all } i, \\ \tilde{\lambda} = \lambda - \mu(Ax^* - y). \end{cases} \quad (3.4)$$

To develop an efficient algorithm which can handle large-scale data sets such as video streams, it is essential to minimize $L_A(x, w_i)$ at a low complexity. An iterative method based on the alternating minimization is employed to minimize $L_A(x, w_i)$ and it is described in the following.

For a given $x$, minimizing $L_A(x, w_i)$ is equivalent to

$$\min_{w_i} \sum_{i=1}^{n} (\|w_i\| - v_i^T(T_i x - w_i) + \frac{\beta_i}{2}\|T_i x - w_i\|_2^2). \quad (3.5)$$

This problem is separable with respect to each $w_i$ and has a closed-form solution [17]:

$$\tilde{w}_i = \begin{cases} \max\left\{|T_i x - \dfrac{v_i}{\beta_i}| - \dfrac{1}{\beta_i}\right\} \mathrm{sgn}(T_i x - \dfrac{v_i}{\beta_i}), \\ \qquad \text{for anisotropic TV,} \\ \max\left\{\|T_i x - \dfrac{v_i}{\beta_i}\|_2 - \dfrac{1}{\beta_i}\right\} \dfrac{(T_i x - v_i/\beta_i)}{\|T_i x - v_i/\beta_i\|_2}, \\ \qquad \text{for isotropic TV.} \end{cases} \quad (3.6)$$

Formula (3.6) can be derived by using 1D and 2D shrinkage.

On the other hand, for a given $w_i$, minimizing $L_A(x, w_i)$ is equivalent to

$$\min_x Q(x) = \min_x \left(\sum_{i=1}^n (-v_i^T(T_i x - w_i) + \dfrac{\beta_i}{2} \|T_i x - w_i\|_2^2) - \lambda^T (Ax - y) + \dfrac{\mu}{2} \|Ax - y\|_2^2\right). \quad (3.7)$$

Clearly, $Q(x)$ is a quadratic function and can be minimized by various iterative methods. However, these methods may be too costly for large-scale data sets and unnecessary since our ultimate goal is to solve (3.1) or (3.2) instead of (3.7). Therefore, a good approximation of the true minimizer of (3.7) should be good enough to guarantee the convergence. Specifically, we take one step of the steepest descent with an aggressive step length and accept it as the approximate solution of (3.7):

$$\tilde{x} = x - \alpha d(x). \quad (3.8)$$

Here, $\alpha$ denotes the step length and $d(x)$ denotes the gradient of quadratic function $Q(x)$ at the previous iteration $x$. As proposed in [17], the step length can be iteratively determined by a nonmonotone line search [26], starting from the initial guess given by the Barzilai-Borwein method [16]. The step length computed like this has been demonstrated to be effective and efficient in practice.

Therefore, the algorithm to minimize the augmented Lagrangian function (3.3) can be described as follows:

**Algorithm 1**. *Initialize starting points and parameters. In particular, initialize $x$ and $w_i$ with initial guess.*

   **while** *not converge*, **do**
     *for the given $x$, compute $\tilde{w}_i$ according to (3.6).*
     *update $w_i \leftarrow \tilde{w}_i$*
     *for the given $w_i$, compute $\tilde{x}$ according to (3.8), by taking one step of steepest descent.*
     *update $x \leftarrow \tilde{x}$*
   **end do**

The above algorithm minimizes the Lagrangian function (3.3) for fixed values of the multipliers $v_i$ and $\lambda$. After Algorithm 1 is performed, these multipliers are updated by using equations (3.4) with $x^*$ and $w_i^*$ replaced by $x$ and $w_i$ computed from Algorithm 1, respectively. After the update of the multipliers, Algorithm 1 is applied again and the multipliers are further updated in another iteration. This defines the iterative process of solving the minimization problem (3.1), and equivalently, (2.2). In order to speed up the convergence of the iteration, a continuation method can be used in which the penalty parameters $\beta_i$ and $\mu$ may be updated from iteration to iteration. For example, at the start of the iteration, the initial values for the penalty parameters are chosen to be much smaller than the target values of the penalty parameters. These values are then increased (or kept the same, but non-decreasing) at each iteration until the target values of $\beta_i$ and $\mu$ are reached. In summary, the overall algorithm is given below:

**Algorithm 2**. *Initialize starting points, multipliers, and penalty parameters.*

   **while** *not converge*, **do**
     *find the minimizers of the augmented Lagrangian function (3.3) by means of Algorithm 1, starting from the previous iterates.*
     *update the multipliers according to (3.4).*
     *choose the new penalty parameters.*
   **end do**

The framework of Algorithm 2 described above can be also used to solve a larger range of minimization problems (2.2) and (2.3) with different regularizations, such as 3D TV, $\ell_1$ under sparsity basis, etc.

## IV. SIMULATION

### A. Simulation using compressive sensing methods

Simulations are performed on two source video clips, Container and News. They are both CIF resolution (352x288) with a frame rate of 30fps. In the simulations, a source video is encoded with compressive measurements as described in Section II.A. A permutated Walsh-Hadamard matrix is used as the measurement matrix. Each cube consists of 8 full CIF frames, i.e., the total number of pixels in each video cube is $n$ =352x288x8=811008. After coding, a number of randomly selected measurements, $m$, is used for reconstruction. The random selection of $m$ measurements simulates a channel of capacity $\dfrac{m}{n}$. That is, it is assumed that a total of $n$ measurements are transmitted, and only $m$ of transmitted measurements are correctly received and used in reconstruction. The PSNR of the reconstructed video is calculated as a function of the percentage of the measurements received ($m/n$).

For each set of correctly received measurements, four different regularizations are used to reconstruct the video including the one proposed in this paper, then the PSNRs of reconstructed videos are compared. To fairly compare the results, reconstruction algorithms for the different models were all implemented similarly to what is described in Section III. The four regularization terms are described below.

***2D TV + pointwise DCT (TV-DCT)***: This is the method proposed in this paper and described in Sections II and III.

***2D TV***: This is the method minimizing the sum of 2D TV of every frame. In other words, the video cube is treated as individual frames, and no temporal relations are explored. Mathematically, let $X$ denote the original 3D cube of $x$ before vectorization, and 2D TV regularization is defined as

$$\Phi(x) = \sum_{i,j,k} \left( \left| X_{i+1,j,k} - X_{i,j,k} \right| + \left| X_{i,j+1,k} - X_{i,j,k} \right| \right).$$

***3D TV***: This is the method minimizing the 3D TV of the whole cube, assuming the sparsity of gradients in both spatial and temporal directions. Mathematically, under the same notation, 3D TV regularization is defined as

$$\Phi(x) = \sum_{i,j,k} \left( \begin{array}{l} \left| X_{i+1,j,k} - X_{i,j,k} \right| + \left| X_{i,j+1,k} - X_{i,j,k} \right| \\ + \left| X_{i,j,k+1} - X_{i,j,k} \right| \end{array} \right).$$

***$\ell_1$ + 3D DCT***: This is the method minimizing the $\ell_1$-norm of the 3D DCT coefficients of the source cube. That is, a 3D DCT transform is performed on the video cube, and the $\ell_1$-norm of the resulting coefficients is minimized. Mathematically, let $DCT_3(x)$ denote the vectorization of 3D DCT of the cube $X$, $\ell_1$ regularization under 3D DCT basis is defined as $\Phi(x) = \left\| DCT_3(x) \right\|_1$.

In all four methods, the same permutated Walsh-Hadamard matrix is used as the sensing matrix, and the received measurements used for reconstruction are randomly chosen. The accuracy of reconstruction, as measured by PSNR of the reconstructed frames with the original frames, is reported in Figures 4-5.

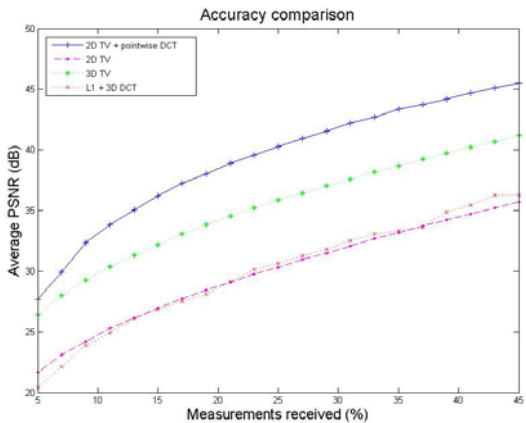

**Figure 4. PSNR for video Container**

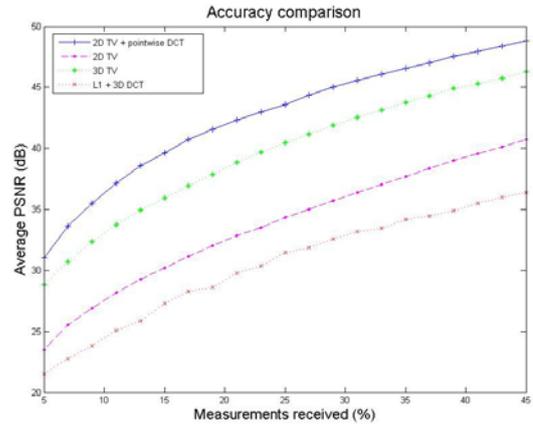

**Figure 5. PSNR for video News**

Two observations can be made from these results. First the compressive sampling methods, 2D TV+pointwise DCT, 2D TV, 3D TV, and $\ell_1$+3D DCT all have the scalability property desired for wireless transmission, namely, the PSNR of the reconstructed video increases progressively with the number of measurements received. This avoids the cliff effect seen with video coding such MPEG2 or H.264.

The second observation is that the TV minimization based methods are superior to the $\ell_1$ minimization of 3D DCT. Among all methods evaluated, the method proposed in this paper is clearly better than the other three.

A typical frame in the recovered video for the Container clip for each of the four tested methods is shown in Figure 6.

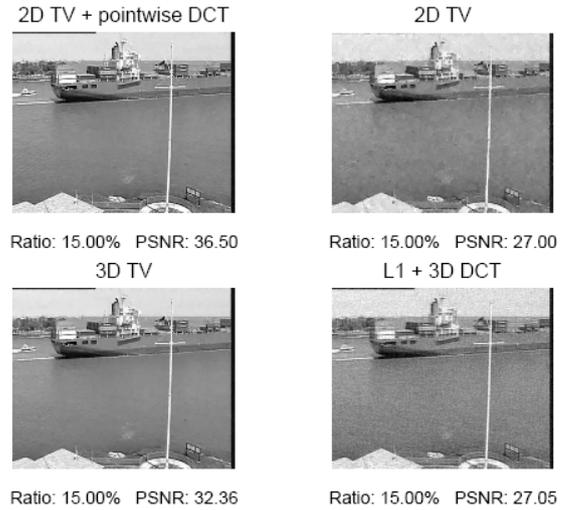

**Figure 6. Frame 4 of recovered video Container**

Figure 6 shows the quality of the reconstructed frame 4 by using different methods. The same number of measurements is used in all methods, and the quality, in terms of PSNR, of the reconstructed video frame for each method is reported. As is evident, the method of this paper, 2DTV+pointwise DCT, has the best PSNR.

Next, we present results regarding the impact of quantization on the performance of the reconstruction using the proposed method. Each received measurement is quantized to a fixed number of bits before they are used for the reconstruction. The PSNR of the reconstructed video is measured as a function of the number of bits in the quantization. The results are reported for different percentages of measurements used in reconstruction, and they are shown in Figures 7-8.

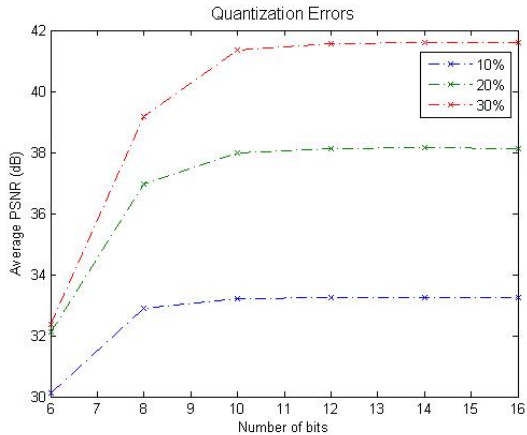

**Figure 7. Quantization impact for Container video**

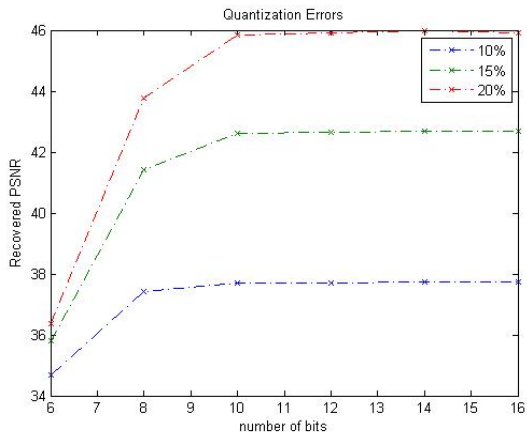

**Figure 8. Quantization impact for News video**

It can be concluded from Figure 7-8 that the reconstruction is not sensitive to the number of bits in the quantization when the number of bits is above 10. For a given quantization level (a given number of bits per measurement), the quality of the video can still be improved by increasing the number of measurements used in reconstruction.

Finally, simulations are performed when channel noise is present. In the simulation, the selected measurements are injected with Gaussian noise, and the resulting noisy measurements are used for reconstruction. In other words, the reconstruction is performed by minimizing (2.3) with $y$ replaced by $\hat{y} = y + n$, where $\hat{y}$ is the received measurements with additive Gaussian noise $n$. The PSNR of the reconstructed video (by using $\hat{y}$) as a function of the carrier to noise level, CNR, is measured. The results are reported for different percentages of measurements used in reconstruction, and they are shown in Figures 9-10.

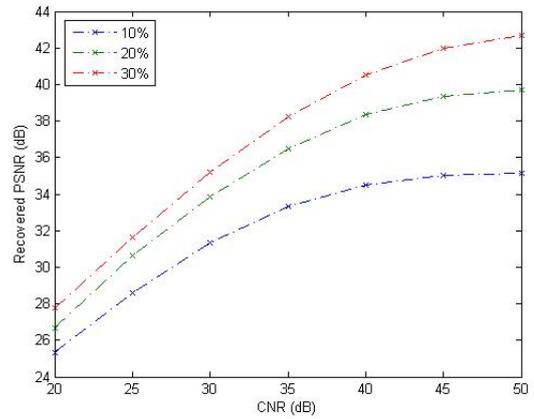

**Figure 9. Impact of additive Gaussian noise for Container video**

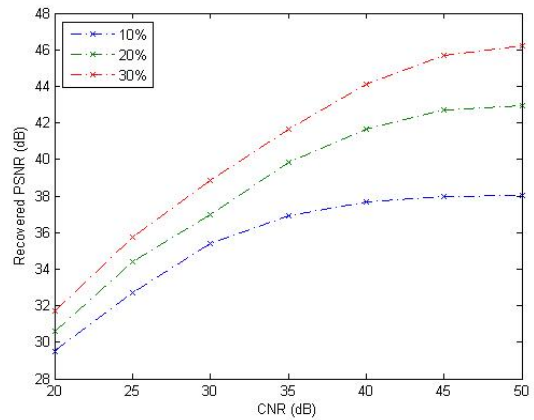

**Figure 10. Impact of additive Gaussian noise for News video**

It is demonstrated in Figure 9-10 that our reconstruction algorithm is reliable when noise is present in the measurements. Furthermore, for a given amount of noise present in the measurements, the quality of video can be improved by using more measurements. This is important for mobile broadcast. The measurements may be broadcast in a method proposed in [8] in which one transmission serves all clients. The amount of channel noise is the sole determination of the quality of the reconstructed video. This further demonstrates the coding method of this paper is scalable with the transmission channel.

*B. Comparison with H.264*

We now present simulation results for H.264 and compare them with the compressive sensing results. In the simulation, we use a broadcast channel model.

**Broadcast erasure channel model**

We consider a broadcast erasure channel (BEC) [30] without feedback. In a BEC model, data is transmitted by a single transmitter, and received by multiple clients. Each broadcast packet is either received correctly by a client or is lost (dropped), in which case an erasure occurs for that client. Each client has its own channel rate $R$, which is the rate of correctly received packets. Different clients may have different channel rate. BEC models have been used to analyze wireless networks such as 802.11, and broadcast systems such as DVB-SH if an uncorrectable packet is dropped and not forwarded to the upper layer.

**H.264 encoding/decoding**

Video sequences are encoded as H.264 streams using a publicly available H.264 codec, FFmpeg [31]. The default encoding parameters are used but the quantization levels are chosen such that the encoded video has a desired PSNR (design PSNR). Specifically, for the News clip, the encoded stream has a bit rate of 440kbps with a design PSNR of 43dB, and for the Container clip, the encoded stream has a bit rate of 490kbps with a design PSNR of 40dB. The rate at which the encoded video achieves the design PSNR is defined as the design rate $R_d$.

After encoding, the H.264 bit streams are modified to emulate the broadcast erasure channel. A number of bits are dropped from the encoded bit streams, and for a client channel rate of $R$, the percentage of the encoded bits correctly received and used in the decoding is equal to $\frac{R}{R_d}$.

Next, the modified bit streams are decoded using the FFmpeg decoder [31] with error concealment turned off. The decoded video is compared with the original video to calculate PSNR.

**Compressive sensing (CS)**

The compressive sensing (CS) method used is the TV-DCT method described previously. Measurements are made by a randomly permutated Walsh-Hadamard matrix. The measurements are then quantized to 10 bits. In BEC model, a measurement is either correctly received or dropped. The client channel rate $R$ represents the number of correctly received measurements per second. The PSNR is calculated by comparing the reconstructed video with the original video.

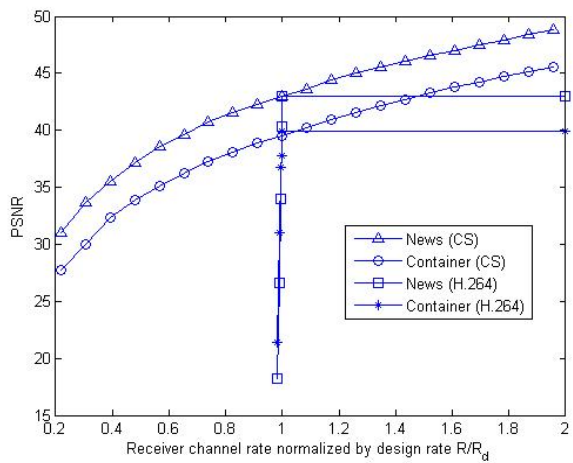

**Figure 11. PSNR vs Normalized client channel rate**

The results from H.264 and CS are compared in Figure 11. The PSNR values of the reconstructed video is plotted against the normalized client channel rate $\frac{R}{R_d}$. The figure clearly shows that standard H.264 codecs exhibit a sharp drop in video quality when the client channel rate is below the design rate $R_d$. Note that the PSNR values for the H.264 streams are less than 20dB when client channel rate is only 2% below the design rate, while the video quality from CS method degrades gracefully. We point out that a PSNR value below 20dB represents severely-degraded video which is non-viewable in practice. In addition, for clients that have higher channel rate than the design rate, the video quality can never get better in H.264 than the design PSNR. Essentially, the H.264 streams are optimized for channel rates in a very narrow range around the design rate.

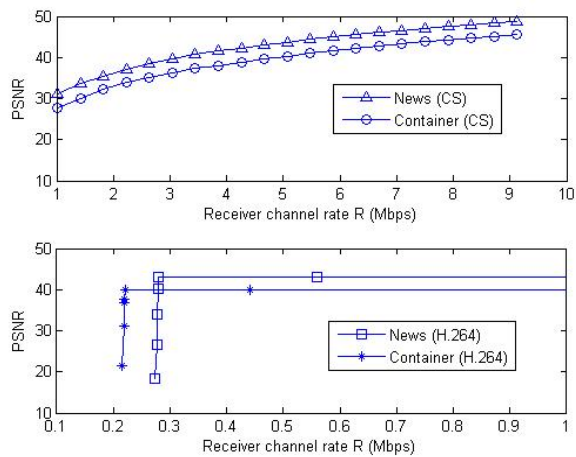

**Figure 12. PSNR vs Client channel rate**

Figure 12 replots the same data as Figure 11 using absolute channel rates. The top plot shows the PSNR values from CS and the bottom plot shows those from H.264. To achieve the same video quality, CS currently requires higher channel rate.

However, this is a somewhat unfair comparison as H.264 has the advantage of both entropy encoding and motion compensation, while CS does not currently use either. Entropy encoding can be easily introduced in CS, and it should have similar impact to CS as it has to H.264. Motion compensation may be introduced during the reconstruction, for example, by an iterative process in which no motion is used in the first pass in the reconstruction, and after the video is reconstructed, a motion estimate is performed to define a trajectory for DCT operation in the next iteration of the reconstruction. Although motion compensation can be used in CS method, it is not clear what its impact is as compared to the impact of motion compensation in H.264. Further study is needed to investigate the true compression ratio of CS method when motion compensation is incorporated.

## V. CONCLUSION

A framework for video coding using compressive sensing is proposed. In this framework, the source video is divided into video cubes, and measurements are made using random matrices. Video is reconstructed by performing the minimization of the framewise total variation of the coefficients from the pixelwise temporal DCT transform. Aiming at the new TV-DCT regularization, an efficient reconstruction algorithm based on TVAL3 is described. This new framework results in a video coding that is suitable for wireless transmission of video. Simulations demonstrate that the proposed method produces videos having higher PSNR than other compressive sampling recovery methods.

The work in this paper is mainly concerned with compressive video coding that is scalable with channel capacity, which is of particular interest in wireless broadcast situations since mobile clients may experience vastly different channel conditions. Further investigation is needed for this framework to become practical in wireless applications. The compression ratio can clearly be improved through the use of entropy encoding, and various schemes to use motion vectors during coding or solely during reconstruction show promise.

## ACKNOWLEDGEMENT

The authors thank Kim Matthews and Gang Huang of Alcatel-Lucent and Wotao Yin of Rice University for insightful discussions. We also thank the reviewers for their valuable comments which improved the paper.